\documentclass[twocolumn, prl, amsmath, amssymb, amsfonts]{revtex4}

\usepackage[T1]{fontenc}
\usepackage{graphicx}
\usepackage{dcolumn}
\usepackage{textcomp}

\usepackage{color}

\renewcommand{\vec}{\boldsymbol}
\newcommand{\db}[2][]{\text{d}^{#1}#2}

\begin{document}
\title{Berezinskii-Kosterlitz-Thouless transition in two-dimensional dipole systems}


\author{A.~Filinov$^{1,2}$}
\author{N.V.~Prokof'ev$^{3,4}$}
\author{M.~Bonitz$^1$}
\affiliation{$^1$ Institut f\"ur Theoretische Physik und Astrophysik, Christian-Albrechts-Universit\"at, Leibnizstr. 15, D-24098 Kiel, Germany\\
$^2$Institute of Spectroscopy of the Russian Academy of Sciences, Troitsk, Russia\\
$^3$Department of Physics, University of Massacusetts, Amherst, Massachusetts 01003, USA\\
$^4$Russian Research Center, Kurchatov Institute, 123182 Moscow, Russia
}

\begin{abstract}
The superfluid to normal fluid transition of dipolar bosons in two dimensions is studied
throughout the whole density range using path integral Monte Carlo simulations and summarized
in the phase diagram. While at low densities, we find good agreement with the universal results
depending only on the scattering length $a_s$, at moderate and high densities, the transition
temperature is strongly affected by interactions and the elementary excitation spectrum.
The results are expected to be of relevance to dipolar atomic and molecular systems and
indirect excitons in quantum wells.
\end{abstract}


\maketitle

Dipolar bosonic systems are of increasing interest for various recent experiments studying the onset of 
superfluidity in nonideal Bose systems and its connection with correlation and quantum degeneracy effects. Examples include dipolar gases, as in recent studies of $^{52}\text{Cr}$ atoms~\cite{Cr}, bosonic molecules, e.g. SrO, RbCs, LiCs, $^{40}$K$^{87}$Rb~\cite{mol}, as well as indirect excitons in semiconductor quantum wells~\cite{exp,ludwig06}.
A number of theoretical and computational studies have addressed the properties of two-dimensional (2D) repulsive dipolar bosons at zero and low temperatures~\cite{mora,Astrak,Buch}. They include the ground-state energy, the structural and coherence
properties, such as the one-body density matrix and the condensate fraction. Quantum Monte Carlo studies at $T=0$ have covered the whole range of coupling strengths up to the crystallization transition. However, the finite-temperature properties, in particular, the superfluid transition temperature $T_c$, remain unexplored.

Previous numerical investigations for 2D homogeneous Bose gases~\cite{denslimit,pilati} have shown that in the dilute (weak coupling) regime, $n a_s^2 \lesssim 10^{-2}$, where $n$ is the density and $a_s$ the $s$-wave scattering length, the exact shape of the interaction potential is irrelevant for $T_c$ which is a function of $n a_s^2$ only~\cite{theor}
\begin{equation}
T_{c}(n)=\frac{2 \pi \hbar^2 n}{m k_B}\frac{1}{\ln(\nu/4\pi) +\ln\ln (1/n a_{s}^2)},
\label{t_kbt}
\end{equation}
where the numerical coefficient is $\nu=380(3)$~\cite{xi}.
However, for moderate and high densities where correlation effects are important no analytical expression is available for $T_c$, calling for investigations
by direct numerical simulations which is the main goal of the present paper. By performing first principle path integral Monte Carlo (PIMC) simulations
we demonstrate that, with increasing interaction strength, the superfluid phase is first stabilized ($T_c$ increases) and then destabilized and vanishes when the system forms a dipolar solid.
We present an explanation of this non-monotonic behavior of $T_c$ suggesting
that it arises from a competition between two types of elementary excitations in the normal fluid
component -- phonons and rotons.

{\bfseries Model and parameters.}
We focus on a pure {\em dipole model} relevant e.g. for various bosonic atoms or molecules and indirect excitons at low densities  where the dipole moment is a free parameter which can be externally controlled, e.g. via an electric field~\cite{buch07,ludwig06}.
The 2D dipole system is described by the Hamiltonian
\begin{equation}
{\hat H} = - \sum_{i=1}^N \frac{\hbar^2\nabla_i^2}{2m} + \frac{1}{2}\sum_{i \ne j} \frac{p^2}{\epsilon_b |\vec{r}_i-\vec{r}_j|^3},
\label{h}
\end{equation}
which is brought to a dimensionless form using the units 
$a=1/\sqrt{n}$ and $E_0=\hbar^2 /m a^2$. The system properties are defined by the dipole coupling $D=p^2/\epsilon_b a^3 E_0$
and the temperature, $T=k_B T/E_0$. 
The thermodynamic equilibrium states of system~(\ref{h}) were sampled by PIMC simulations with the worm algorithm~\cite{prokof}.
We studied the system (\ref{h}) from weak to strong coupling, $D=0.01\ldots 20$. 
In agreement with Refs.~\cite{mora,Astrak,Buch} we
observe formation of a crystalline phase at $D \simeq 18$.

\begin{figure}
\vspace{-1.95cm}
\includegraphics[width=0.45 \textwidth]{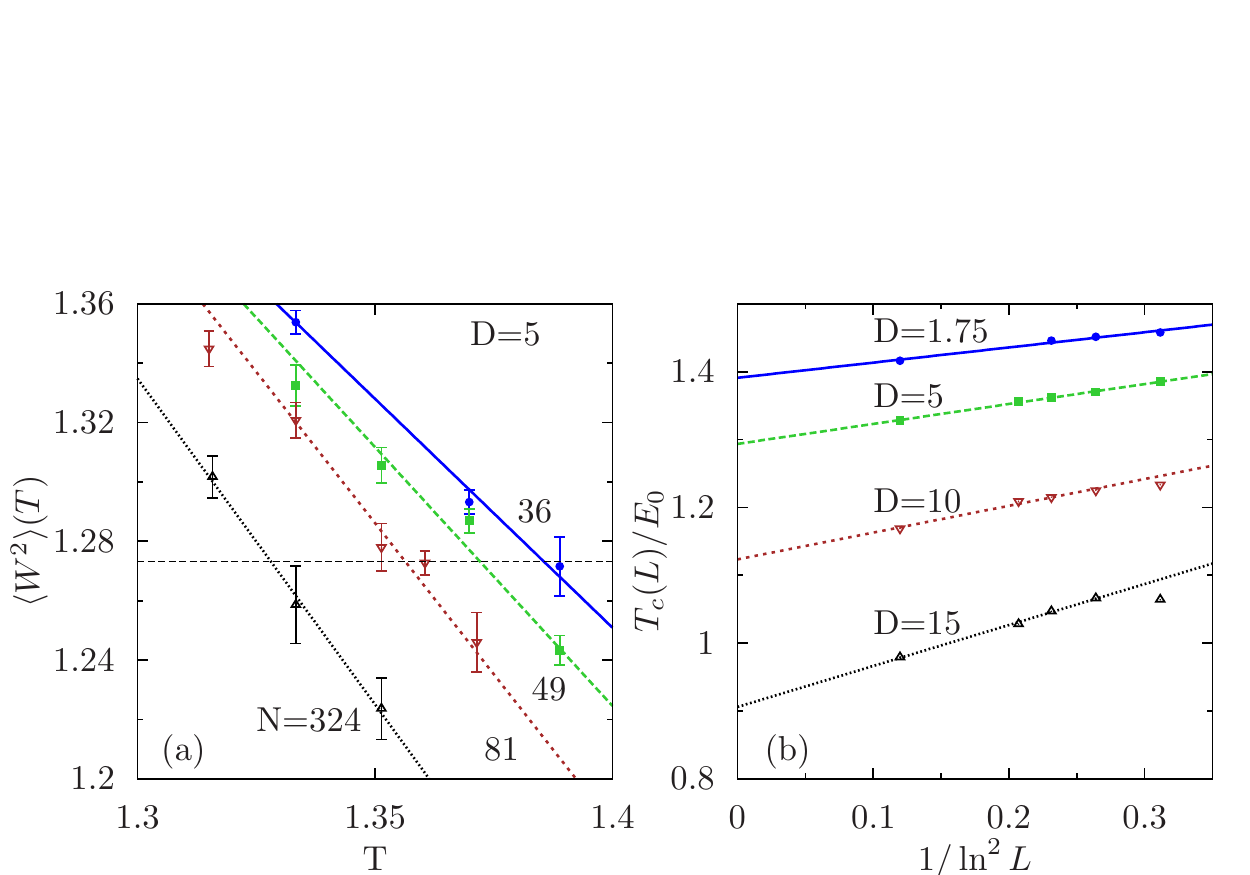}
\vspace{-0.5cm}
\caption{(a) Winding number $\langle W^2 \rangle(T)$ for different system sizes: $N=36,49,81,324$. Dipole coupling $D=5$. Horizontal dashed line:
$ W^2(T_c)=4/\pi$. (b) System size dependence of the critical temperature for several coupling strengths $D$.}
\label{fig2}
\end{figure}

{\bfseries Superfluid transition and phase diagram.}
In 2D the superfluid-normal phase transition occurs at a finite temperature $T_c$ and follows the Berezinskii-Kosterlitz-Thouless scenario (BKT) induced by interaction effects~\cite{bkt}. To obtain a reliable result for $T_c$ in the thermodynamic limit, $T_c(\infty)$, from simulations of a finite system of size $L=\sqrt{N}$, we apply a finite-size scaling analysis to $T_c(L)$.
We assume the essential singularity~\cite{kt} of the correlation length $\xi(T)\sim e^{a t^{-1/2}}, t=(T/T_c-1)|_{T\rightarrow T_c}$, with $a$ being a non-universal temperature-density dependent scaling factor. Near $T_c$ the role of $\xi$ is taken over by $L$, leading to
($b$ is a constant, cf. Fig.~\ref{fig2}.b)
\begin{eqnarray}
T_c(L)=T_c(\infty)+ \frac{b}{\ln^2(L)}, \qquad
n_s(L,T_c)=\frac{2m k_B}{\pi \hbar^2} T_c,
\label{tc_l}
\end{eqnarray}
where $T_c(L)$ is determined by the scenario of the universal jump
of the superfluid fraction (second equation) \cite{superf}.
Here, the superfluid density $n_s$ is obtained via the {\em winding number} estimator~\cite{cep99},
$n_s(L,T)=m k_B T \langle \mathbf{W}^{2} \rangle(T,L) /2\hbar^2$
which is directly evaluated by PIMC simulations. Combining this with
Eq.~(\ref{tc_l}),
$T_c(L)$ is determined by the condition $\langle \mathbf{W}^2 \rangle(T_c(L),L)=4/\pi$.

In Fig.~\ref{fig2} we show the temperature dependence and the finite-size scaling of $\langle W^2 \rangle(T,L)$ and $T_c(L)$.
We observe that $T_c(L)$  shifts systematically with $L$ to lower values (Fig.~\ref{fig2}a).
The extrapolation to the thermodynamic limit, $T_c(\infty)$, fitting the simulation data by Eq.~(\ref{tc_l}), is reported in Fig.~\ref{fig2}b.
For strong coupling ($D\geq 10$) the finite-size corrections to $T_c$ in Eq.~(\ref{tc_l}) become important, therefore, we excluded from the fit the smallest system ($N=36$).

\begin{figure}[h]
\centering\hspace{0cm}\includegraphics[width=0.41 \textwidth]{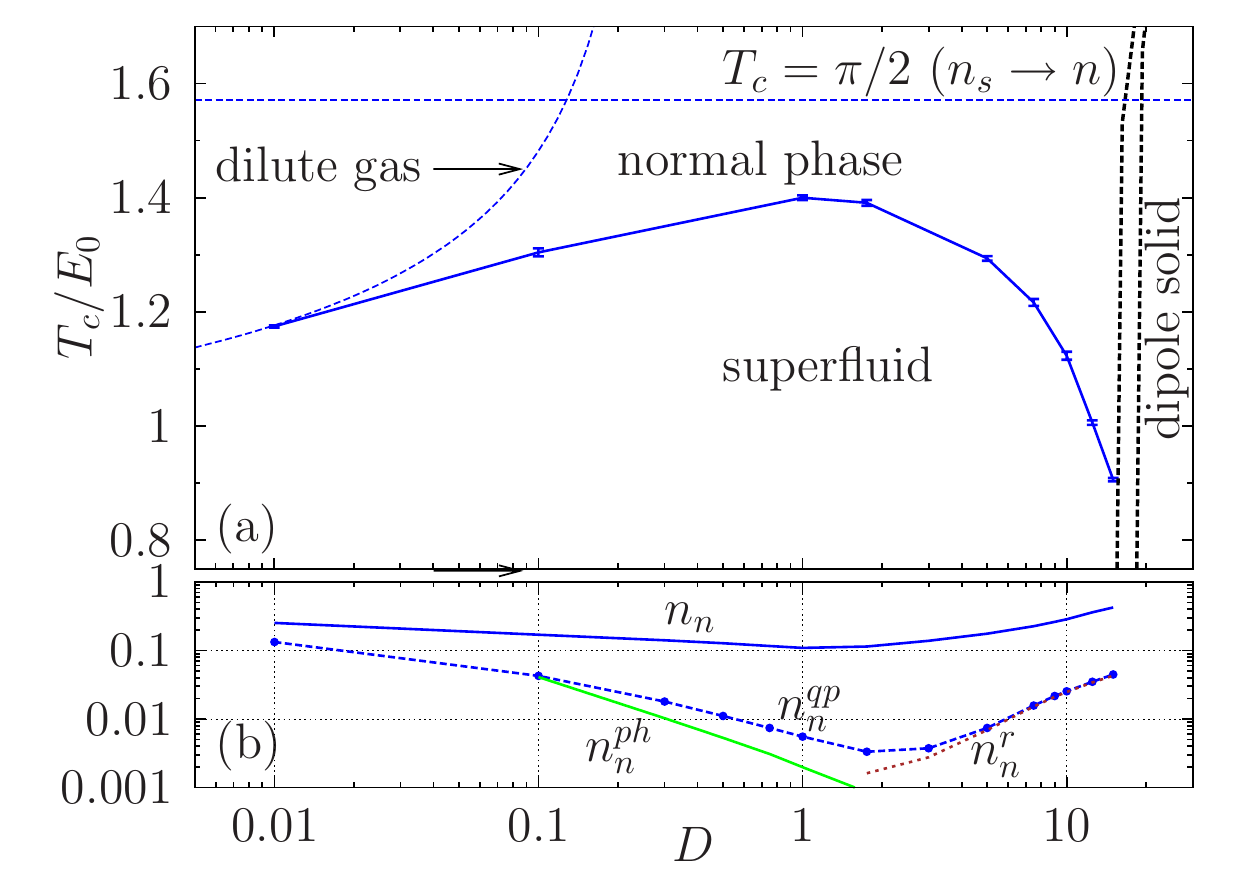}
\vspace{-0.3cm}
\caption{(a): Phase diagram in the $T$-$D$ plane. The system is superfluid below the solid (blue) line.
The two vertical dashed lines bound the gas-solid hysteresis region to the right of which the system is in a dipole crystal phase \cite{kalia,mora,Astrak,Buch}.
The horizontal dashed line is the upper bound for $T_c$ obtained by replacing
$n_s \rightarrow n$.
The dotted line denotes the estimate for $T_c$ according to Eq.~(\ref{t_kbt}).
(b): $D$-dependence of the normal density $n_n$ (PIMC result), the quasiparticle contribution $n_n^{\text{QP}}$, Eq.~(\ref{Land}) and
the phonon and roton contributions, Eqs.~(\ref{n_ph}) and (\ref{n_r}), respectively, for $T=T_c$.
The difference between $n_n$ and $n_n^{\text{QP}}$ equals $n_n^{\text{v}}$ and is due to
vortices, cf. table~\ref{tab1}.}
\label{fig_phase}
\end{figure}

Using the extrapolated data, $T_c(\infty,D)$, we construct the phase diagram, cf.~Fig.~\ref{fig_phase} and Table~\ref{tab1}.
At small coupling, $D \lesssim 0.01$, our results are well reproduced by the asymptotic expression~(\ref{t_kbt}), i.e. here, details of the interaction potential are not important.  However, the validity range of Eq.~(\ref{t_kbt}) is limited to very low densities, $n a_s^2\lesssim 10^{-3}$ [computing
the scattering length from the solution of the Schr\"odinger equation for 2D dipoles gives the relation $n a_{s}^2\approx 10.05 D^2$].
This density is an order of magnitude smaller than for a 2D gas of hard disks~\cite{pilati} indicating that the long-range character of the dipole interaction causes substantially earlier deviations from the dilute gas limit.

Now we analyze, in more detail, the change of $T_c$ with coupling, Fig.~\ref{fig_phase}.
For small $D$, $T_c$ monotonically increases and reaches a maximum around $1<D<2$ whereas, for larger couplings, $T_c$ monotonically decreases again until the system freezes into a non-superfluid dipole crystal. [This is preceded by a narrow hysteresis region ($15  < D  < 18$), shown by the vertical dashed lines, where a bubble-type structure is expected~\cite{spivak} which, however, is beyond the scope of the present paper.]
%
In the following, we 
demonstrate that the non-monotonic behavior of $T_c(D)$ can be explained by the excitation spectrum.

%
%

\begin{figure}[h]
\vspace{-1.95cm}
\includegraphics[width=0.45\textwidth]{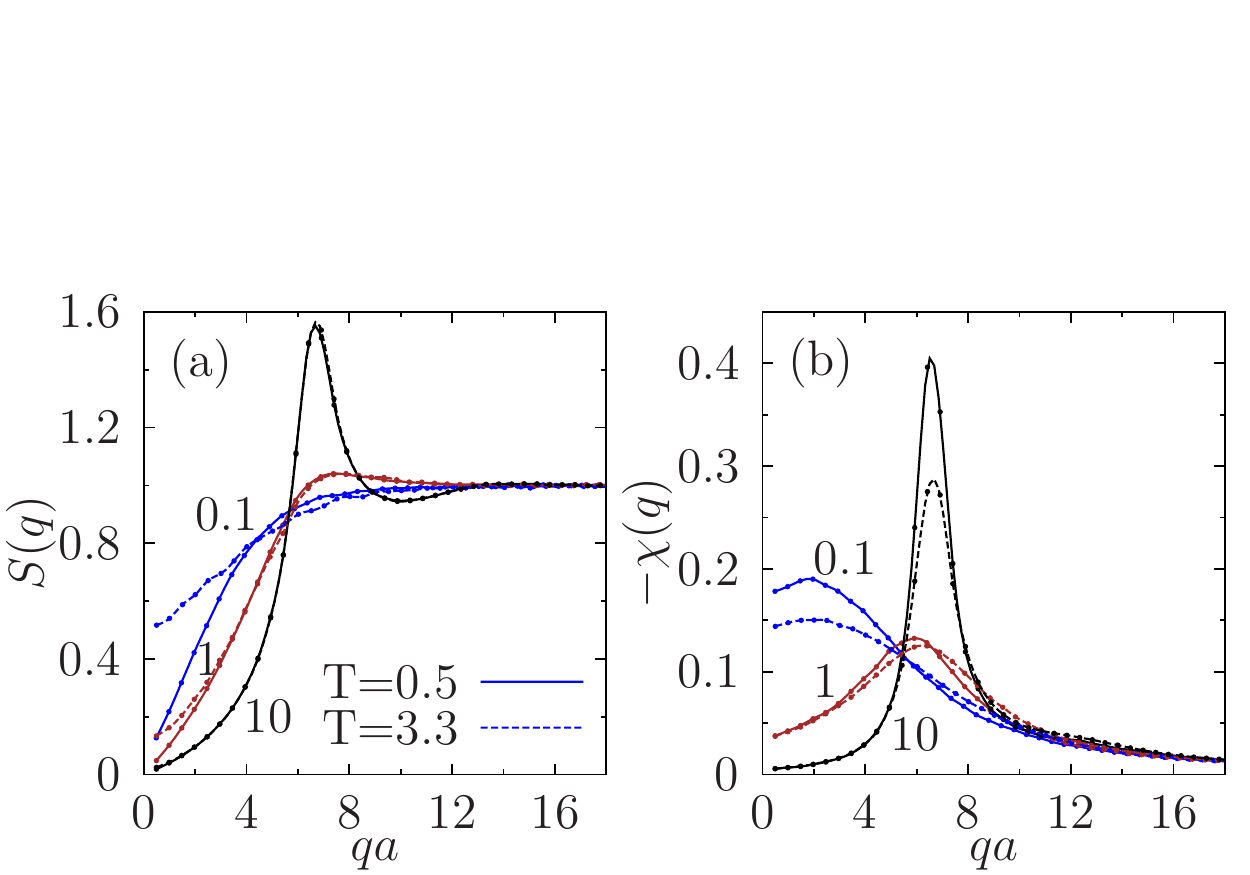}
\vspace{-0.4cm}
\caption{(a) Static structure factor and (b) density response function at $T=0.5$ (solid lines) and $T=3.3$ (dashed lines) for three couplings, $D=0.1, 1, 10$ (numbers in the figure).}
\label{s_stat_old}
\end{figure}

\begin{figure}[h]
\begin{center}
\hspace{-0.7cm}\includegraphics[width=0.55\textwidth]{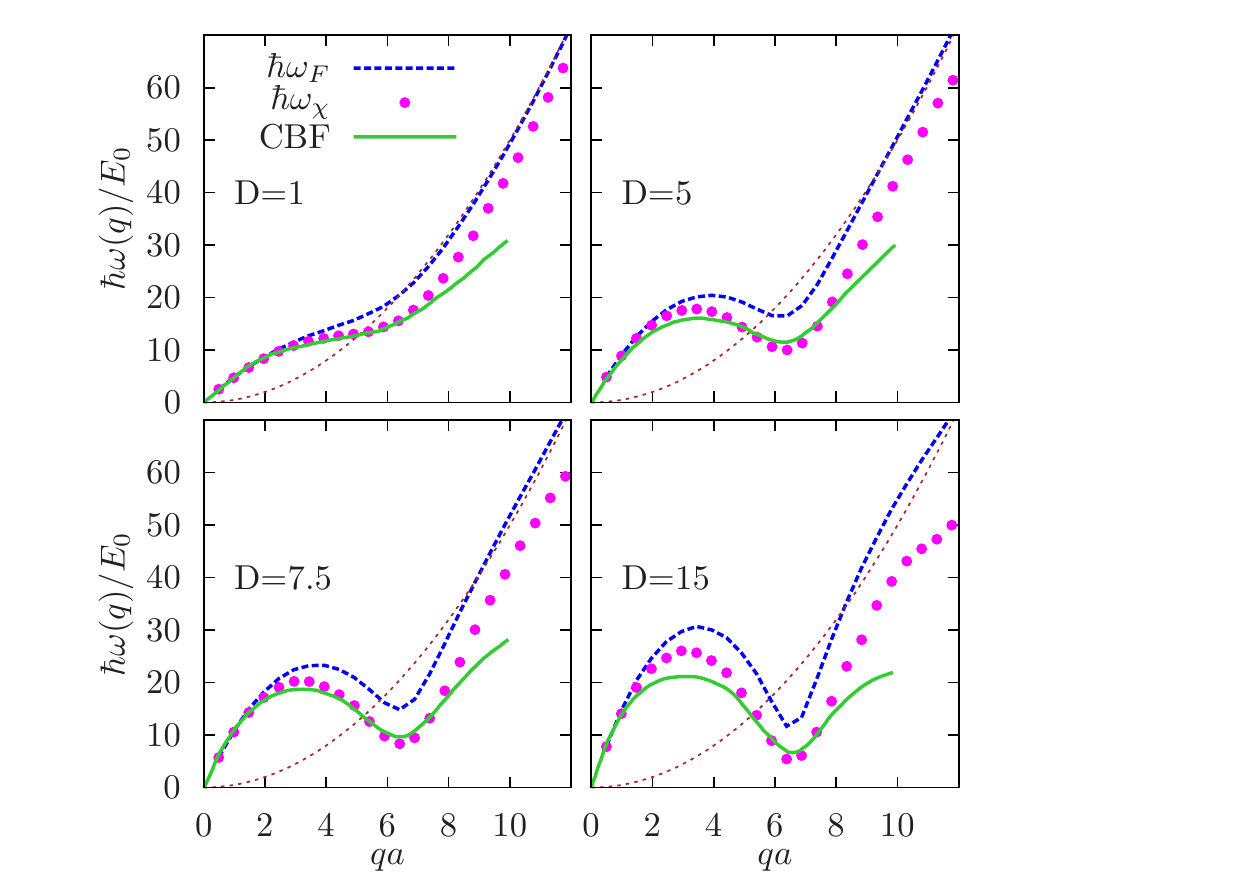}
\vspace{-0.7cm}
\caption{Excitation spectrum $\varepsilon_q=\hbar \omega(q,T)$ for $T=0.5$ and different couplings $D$.
PIMC results, $\omega_F$, Eqs.~(\ref{fe}), and $\omega_{\chi}$, Eq.~(\ref{w_chi}), are compared with the CBF spectra ($D=1,4,8,16$), Ref.~\cite{mazz} and
$(\hbar q)^2/2m$ (short dashed lines).}
\label{s_stat}
\end{center}
\end{figure}

{\bfseries Excitation spectrum.}
The excitation spectrum $\omega(q)$ can be approximated from above through the finite temperature generalization of the Feynman relation~\cite{FS}
\begin{equation}
\omega(q) \leq \omega_F(q), \; \omega_F(q) \tanh \left[\frac{\hbar \omega_F(q)}{2 T} \right]=\frac{\hbar q^2}{2m S(q,T)},
\label{fe}
\end{equation}
where $S(q,T)$ is the static structure factor which we computed in the PIMC simulations for a broad range of temperatures, $0.5\leq T \leq 3.3$, below and above $T_c$, cf. Fig.~\ref{s_stat_old}a. As $D$ approaches the crystallization point, a sharp
peak develops near the wave number $q_0$ corresponding to the mean interparticle distance, $q_0 a=2 \pi$.
Note that, while $S(q,T)$ shows some $T$-dependence for $qa  <3$,
the Feynman  spectrum, $\omega_F(q,T)$, stays almost unchanged in a broad temperature interval, $T\lesssim 3.3$, and is close to the ground state result~\cite{Astrak,helium-spectrum}. Therefore, the spectra shown in Fig.~\ref{s_stat} for $T=0.5$
are representative for the low-temperature behavior.

In the long wavelength limit, $qa \rightarrow 0$, $\omega_F$ becomes exact yielding a linear dispersion, $\omega_F(q)= c_s q$, cf. Fig.~\ref{s_stat}a
which is in agreement with the result for classical 2D dipoles~\cite{kach,kalm}.
Our results for the sound speed, $c_{s}(T)=\omega_F(q,T)/q|_{q\rightarrow 0}$, extracted from the data for $N=324$ particles
are summarized in Tab.~\ref{tab1} and agree within $4\%$ with the ground state values of Ref.~\cite{Astrak}. 

A significant improvement of the spectrum is achieved by using a sum-rule approach~\cite{string,moroni}
by combining the PIMC results for $S(q,T)$ and the static density response $\chi(q,T)$, yielding a rigorous upper bound
\begin{align}\label{w_chi}
&\hbar \omega(q) \leq \hbar \omega_{\chi}(q,T) = 2 n S(q,T)/\chi(q,T),\\
& \chi(q)=-\int_0^{\beta} F(q,\tau) \, \db \tau, \; F(q,\tau)=\frac{1}{N}\left\langle\rho_q(\tau) \rho_q(0)\right\rangle,
\nonumber
\end{align}
where $\chi(q)$ is obtained from the imaginary-time density-density correlation function $F(q,\tau)$
which has been directly evaluated in PIMC simulations.
With the increase of $D$ (Fig.~\ref{s_stat_old}b) the response function sharpens and shifts continuously towards $q_0$.

In Fig.~\ref{s_stat} we show $\omega_{\chi}$, Eq.~(\ref{w_chi}), together with the Feynman spectrum and the correlated basis functions (CBF) result~\cite{mazz} at four dipole couplings. All three approximations show the same general trend which resembles superfluid helium: with increasing coupling the spectrum develops a roton minimum at finite $q\approx q_0$ which becomes deeper with increasing $D$. While for $qa \lesssim 1.5$ (sound range) all approaches are in quantitative agreement, for $qa>2$ the Feynman approximation becomes inaccurate. Its error increases with $D$ and exceeds $100\%$ at the crystallization point for $\omega$
at the minimum. The PIMC result $\omega_{\chi}(q)$ agrees suprisingly well with $\omega_{\text{CBF}}(q)$. Our simulations predict a deeper minimum $\omega(q_{0})$ and are expected to be more accurate here. 
Further, for $q\gtrsim 7.5$, the upper bound  $\omega_{\chi}(q)$ approaches a free-particle spectrum (similar to $\omega_F$), except for
$D \gtrsim 15$, whereas CBF,
at strong coupling ($D=15$), shows the onset of a plateau.
In analogy with superfluid $^4$He a plateau might be expected at twice the roton minimum energy but it appears that all schemes
violate this threshold and this question requires additional research.

We now return to the phase diagram.
$T_c(D)$ in Eq.~(\ref{tc_l}),
%
is related to the normal density by $n_s(T)=n-n_n(T)$ which
for non-interacting quasiparticles is determined by the excitation spectrum $\varepsilon_p=\hbar \omega_{\chi}(p;D,T)$ via the Landau formula~\cite{Khal}
\begin{align}
n^{\text{QP}}_n(T)=-\frac{1}{2}\frac{1}{(2\pi \hbar)^2} \int_0^{\infty} d^2 p\, p^2 \frac{d n_B}{d \varepsilon_p},\;
\label{Land}
\end{align}
where
$n_B=[e^{\beta \varepsilon_p}-1]^{-1}$ is the Bose distribution function, and
we used $m=1$. The superscript $QP$ denotes the dilute quasiparticle approximation.

At low temperatures the main contribution to $n^{\text{QP}}_n$, Eq.~(\ref{Land}), comes from the lowest energy quasiparticles -- phonons -- with the dispersion $\varepsilon^{ph}_p=c p$, and rotons, $\varepsilon^{r}_p=\Delta+(p-p_0)^2/2 \mu$, with the approximate result~\cite{Khal} in 2D
\begin{eqnarray}
n_{n}^{\text{ph}} &=& \frac{3 \bar{\varepsilon}_{\text{ph}}}{2 c^2}=\frac{3 \zeta(3)}{2 \pi} \frac{(k_B T)^3}{\hbar^2 c^4}=0.574 \frac{(k_B T)^3}{\hbar^2 c^4}, \label{n_ph} \\
n_{n}^{r} &\approx & \frac{\beta p_0^2}{2} n_r, \; n_r(T)=\frac{p_0}{2 \pi \hbar^2} e^{-\beta \Delta} (2 \pi \mu k_B T)^{1/2}.\label{n_r}
\end{eqnarray}
The free parameters in the dispersions $\varepsilon^{ph}_p$ and $\varepsilon^{r}_p$ are determined from a fit to the simulation result
$\hbar \omega_{\chi}(p)$. The values for $c$ ($c\approx c_s$) and the roton gap $\Delta$ are summarized in Table~\ref{tab1}.
We note that $\Delta$ in the liquid phase decreases exponentially with the dipole coupling:
$\Delta(D)/E_0=a_1 \exp(-a_2 D -a_3 D^2)$,
with the best fit parameters: $a_1=15.11(5)$,  $a_2=0.088(2)$, $a_3=-0.00120(8)$,
whereas at the crystallization point we find $\Delta(D=18)/E_0=4.57$.

Consider now the $D$-dependence of $n_{n}^{\text{ph}}$ and $n_{n}^{r}$, cf. lower part of Fig.~\ref{fig_phase}.
At weak coupling, $D \lesssim 1$, there is no roton minimum in the spectrum, cf. Fig.~\ref{s_stat}, $D=1$, and $n_n^r=0$.
At the same time there
is a strong phonon contribution, $n_n^{ph}$, which monotonically decreases with $D$ since, at a fixed temperature, the sound speed increases with coupling, $c_s \sim D^{1/2}$, cf. Tab.~\ref{tab1}.
As soon as the spectrum contains a roton minimum, i.e. for $D \gtrsim 1.5$, cf. Fig.~\ref{s_stat}, phonon excitations are practically
negligible and the rotons dominate which is due to the larger density of states ($\sim q dq$). The roton density $n_n^r$ monotonically increases with $D$, cf. Fig.~\ref{fig_phase}.b due to the reduction of the roton gap, 
cf. Tab.~\ref{tab1}. These competing trends of phonon and roton excitations give rise to a minimum in the sum $n_n^{ph}(D)+n_n^r(D)$ at a coupling around
$D=1.5$. Note that this sum of the approximate contributions (\ref{n_ph}), (\ref{n_r}) rather well reproduces the full quasiparticle density $n^{\text{QP}}_n(D)$, Eq.~(\ref{Land}).

Most interestingly, the position of the minimum is very close to the maximum of the superfluid transition temperature $T_c$ which is observed in the range $D\simeq 1-1.75$, and the $T_c(D)$ appears to follow just the opposite behavior of the curve $n^{\text{QP}}_n(D)$.
In fact, this opposite trend could be expected from Eq.~(\ref{tc_l}) stating that $T_c \sim n-n_n$ where $n_n(T_c)$ is the full normal density at the critical point. However, there is no obvious reason why $n_n(D)$ should follow $n^{\text{QP}}_n$ in the case of strongly interacting bosons. Our simulations allow us to directly compare the
exact value, $n_n$ to $n^{\text{QP}}_n$. As can be seen in Fig.~\ref{fig_phase}.b there is a dramatic difference between the two indicating the existence of an additional contribution $n_n^{\rm v}$ to the normal density, i.e. $n_n = n^{\text{QP}}_n + n_n^{\rm v}$, which is by far dominant in the region around the maximum of $T_c$ exceeding $n_n^{\text{QP}}$ by as much as a factor thirty.

The physical origin of $n_n^{\rm v}$ are interaction effects and vortices missing in the model (\ref{Land}). Large scale vortices are a
key in the BKT theory and they are known to substantially reduce the true macroscopic superfluid density $n_s$ compared to its local value  
$n_s^{loc}(T)$. It has been shown that this suppression is similar to screening ~\cite{bkt,Minn} characterized by an effective dielectric function, $n_s=n_s^{loc}/\epsilon$. For the present system, it is reasonable to identify
$n_s^{loc} \rightarrow n - n^{\text{QP}}_n$ which relates $\epsilon$ to $n_n^{\rm v}$ according to
$n_n^{\rm v} = (n-n^{\text{QP}}_n)(1-\epsilon^{-1})$. Our simulations allow us to directly compute the vortex density and $\epsilon(D)$ for all couplings, the results are included in table~\ref{tab1}. Notice that $n_n^{\rm v}$ and $\epsilon$ rapidly increase for strong coupling. However, for $D\lesssim 3$, i.e. in the range where $T_c(D)$ reaches its maximum they are almost $D$-independent. This behavior is constrast to the one of $n_n^{ph}(D)$ and $n_n^r(D)$. Thus we may conclude that - despite the small absolute value of $n_n^{ph}(D)$ and $n_n^r(D)$ compared to $n_n^{\rm v}$, it is the phonon and roton excitations which are responsible for the shape of the phase boundary $T_c(D)$ and for the stabilization of the superfluid phase around $D\simeq 1-1.75$.

\begin{table}[h]
\caption{Coupling parameter dependence of the superfluid transition temperature $T_{c}$, the superfluid fraction $\gamma_s(T_c)=2m T_c/\pi \hbar^2 n$, the sound speed $c_s$ [in units of the dipole frequency,
[$\omega_D \bar{a}$], $\omega_D^2=2 \pi p^2 n /(m \bar{a}^3)$, $\bar{a}=(\pi n)^{-1/2}$], the vortex density $n_n^{\text{v}}(T_c)$ and  the effective dielectric function, $\epsilon(T_c)$.}
\label{tab1}
\begin{tabular}{c|c|c|c|c|c|c}
\hline
\hline
$D$ & $T_{c} [E_0]$& $\gamma_s(T_c)$ & $c_s$ & $\Delta[E_0]$ & $n_n^{\text{v}}(T_c)$ & $\epsilon(T_c)$\\
\hline
\hline
0.01&1.174(2)  &0.75 & 3.60 & --& 0.12 &1.16\\
0.1&1.304(7) &0.83 &2.23 & --& 0.13&1.15\\
1&1.400(4) &0.89&1.59& -- & 0.10&1.12\\
1.75&1.391(5) &0.88& 1.48&13.02 & 0.11&1.13\\
3&1.353(5) &0.86& 1.41&11.73& 0.13&1.15\\
5&1.294(4) &0.82&1.36& 10.00& 0.17&1.20\\
7.5&1.216(6) &0.77& 1.31& 8.35&0.21&1.27\\
10&1.123(7) &0.71& 1.30& 7.12&0.26&1.36\\
12.5& 1.006(4) & 0.64& 1.28& 6.06& 0.32&1.51\\
15&0.906(3)&0.58& 1.28& 5.27&0.38&1.66\\
\hline
\hline
\end{tabular}
\end{table}

In conclusion, the finite-temperature phase diagram of a 2D dipole system has been investigated by first-principle PIMC simulations over the entire coupling regime. We found that 
the superfluid density at 
$T_c$ does not exceed $90\%$ and drops to about $58 \%$ near the crystallization point.
The normal density is dominated by vortices and contains a small fraction of phonons and rotons. Yet the competition of the latter two is responsible for the observed non-monotonic behavior of $T_c(D)$.
Furthermore, an upper bound for the single-particle spectrum has been computed which significantly improves the result of the Feynman approximation. We expect that our predictions are of direct relevance for experiments with  atomic and molecular dipole systems as well as for indirect excitons in
semiconductor quantum wells.

This work is support by the Deutsche Forschungsgemeinschaft via grant FI 1252/1.


\begin{thebibliography}{26}
\bibitem{helium-spectrum} A similar weak temperature dependence of the excitation spectrum is also observe for liquid helium up to $T\sim 2 T_{\lambda}$~\cite{Sven}.
\end{thebibliography}
\end{document}